\documentclass[twoside]{dis04}
\usepackage{amsmath}
\def\runauthor{J.P. Lansberg et al}
\def\shorttitle{A Simple Model for Pion GPD}

\begin{document}

\title{A Simple Model for Generalised Parton Distributions of the Pion}

\author{J. P. Lansberg$^\dagger$, F. Bissey$^*$, J. R. Cudell, J. Cugnon, P. Stassart}

\address{D\'epartement de  Physique, B\^at. B5a, Universit\'e de  Li\`ege, 
Sart Tilman, B-4000 Li\`ege 1, Belgium\\
$^*$Institute of Fundamental Sciences, Massey University, Private Bag 11 222, 
Palmerston North, New Zealand\\
$^\dagger$E-mail: JPH.Lansberg@ulg.ac.be }

\maketitle

\abstracts{We present here an extension of our model for the pion, 
which we used previously to calculate its diagonal 
structure function, to the off-forward case.
The imaginary part of the off-forward $\gamma^{\star} \pi \rightarrow
\gamma^{\star}\pi$ scattering amplitude is evaluated in the chiral limit ($m_\pi=0$) 
and related to the twist-two and twist-three
generalised parton distributions ${H}$, ${H}^3$, $\tilde {H}^3$. 
Non-perturbative effects, linked to the size of the pion and still preserving 
gauge invariance, are included. 
Remarkable new relations 
between ${H}$, ${H}^3$ and $\tilde {H}^3$
are obtained and discussed.
}

\section{Introduction} 

Structure functions are  useful tools to understand the structure of hadrons. At large $Q^2$, 
they are related to  parton distributions. Although their $Q^2$-evolution  is consistent with 
perturbative QCD, their bulk properties come from nonperturbative effects. The latter are 
often treated by low-energy models, such as NJL, which establish a connection with the low 
$Q^2$ physics. There has been extensive work on diagonal
distributions along these lines (see Ref.~\cite{ours} and references therein for the pion case). 

The interest has now turned to the off-diagonal case~\cite{our2,Belitsky:2000vk,GPD_pion}. 
For the latter, the off-forward structure functions are related to the 
off-forward $\gamma^*$-hadron amplitude and appear as convolutions of generalised parton 
distributions. These carry information about correlations between partons. In order to 
illustrate the properties of these quantities, we undertook to calculate them in the 
case of  the pion. In Ref.~\cite{ours}, we first calculated the forward amplitude 
and the quark distribution in a simple model, in which the pion field is coupled 
to (constituent) quark fields through a $\gamma^5$ vertex. Furthermore, pion size effects 
are introduced through a gauge-invariant procedure by requiring that the squared 
relative momentum of the quarks inside the pion is smaller than a cut-off value. The most 
remarkable result of this investigation is that the momentum fraction carried   by the quarks 
is  smaller than  one, although gluonic degrees of freedom are not included. Here, we report 
on the extension of our model to the off-diagonal case~\cite{our2}.

In the following, we  calculate the imaginary part of the
off-forward photon-pion scattering amplitude,
and of the structure functions $F_1,\dots,F_5$, related to 
the five independent
tensor structures in the scattering amplitude, and we
discuss their behaviour. We relate them to vector and axial vector form factors
and to the twist-two 
and twist-three generalised parton distributions (GPD's)  
${H}$, ${H}^3$ and $\tilde {H}^3$. We shall show that, within 
our model and in the high-$Q^2$ limit, the non-diagonal structure functions 
$F_3$ and $F_4$ are related to $F_1$, while $F_5$ happens to be a higher twist.
These results lead to new relations for the GPD's in the neutral pion case.

\section{TENSORIAL STRUCTURE OF THE   $\gamma^{\star} \pi \rightarrow 
\gamma^{\star}\pi$ AMPLITUDE }

\begin{figure}[!thb]
\centering
\mbox{\psfig{figure=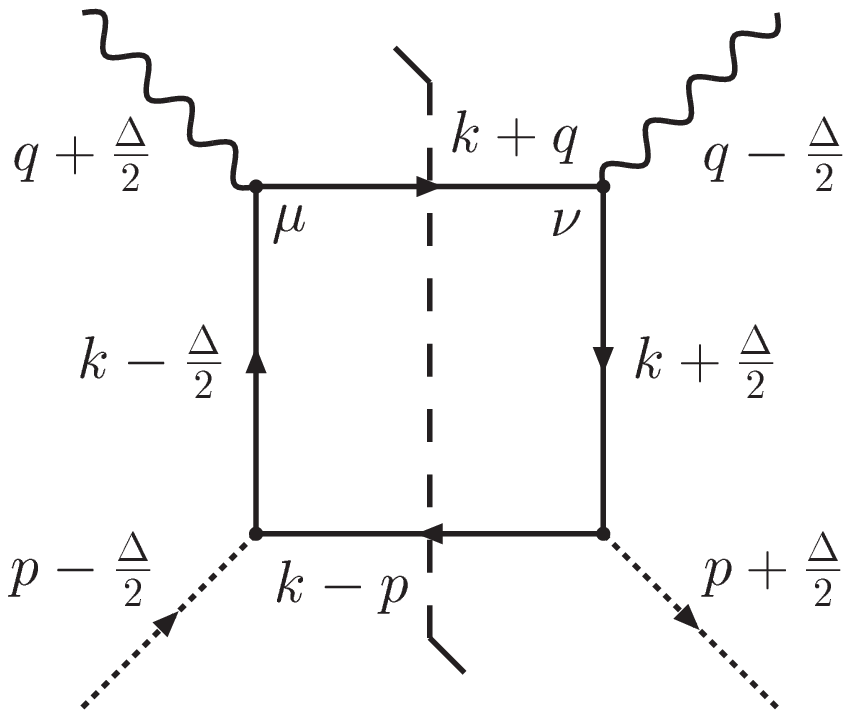,height=1.5cm,}\quad
      \psfig{figure=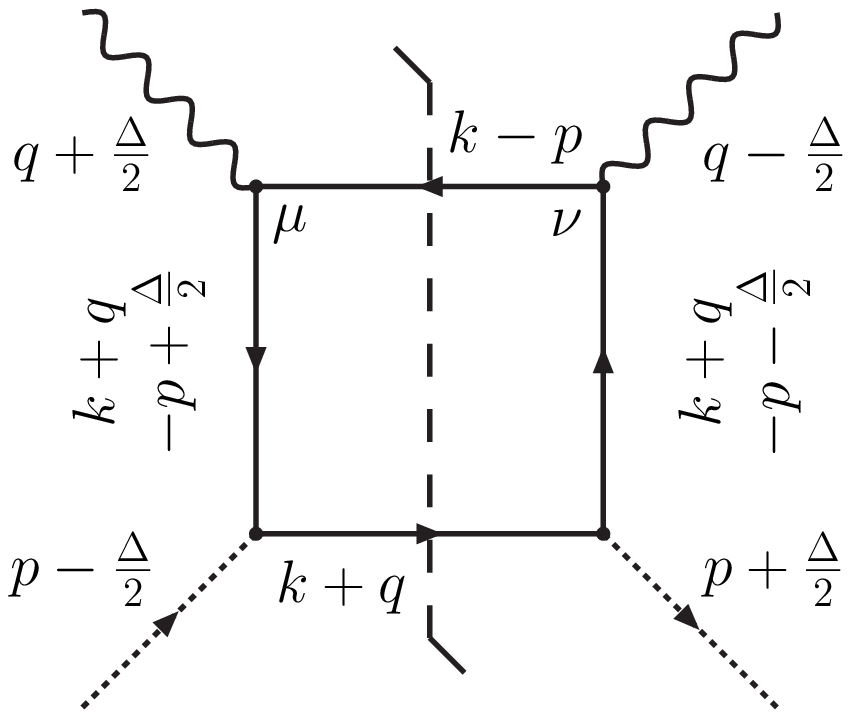,height=1.5cm}
\psfig{figure=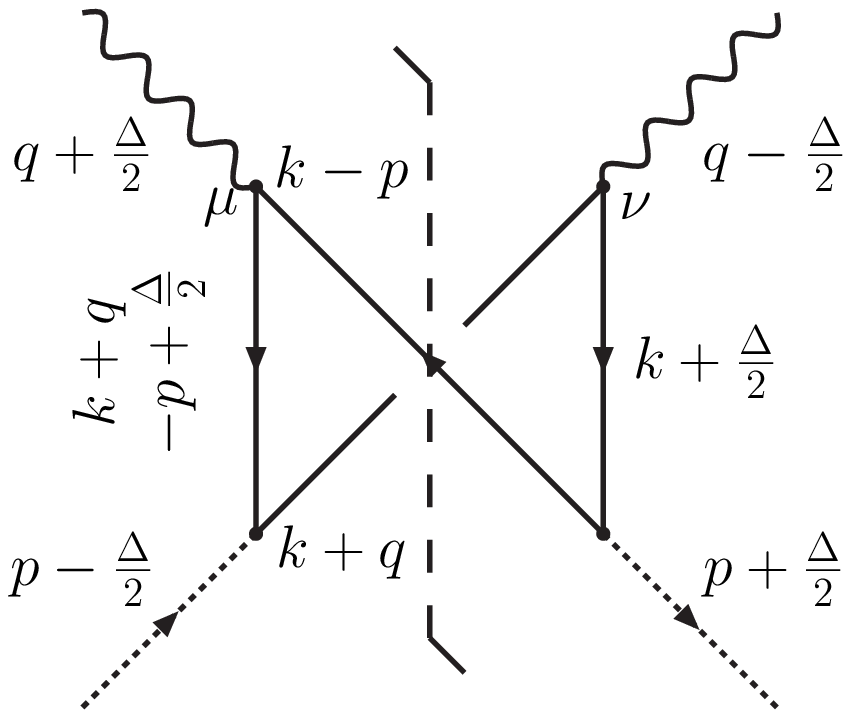,height=1.5cm}\quad
      \psfig{figure=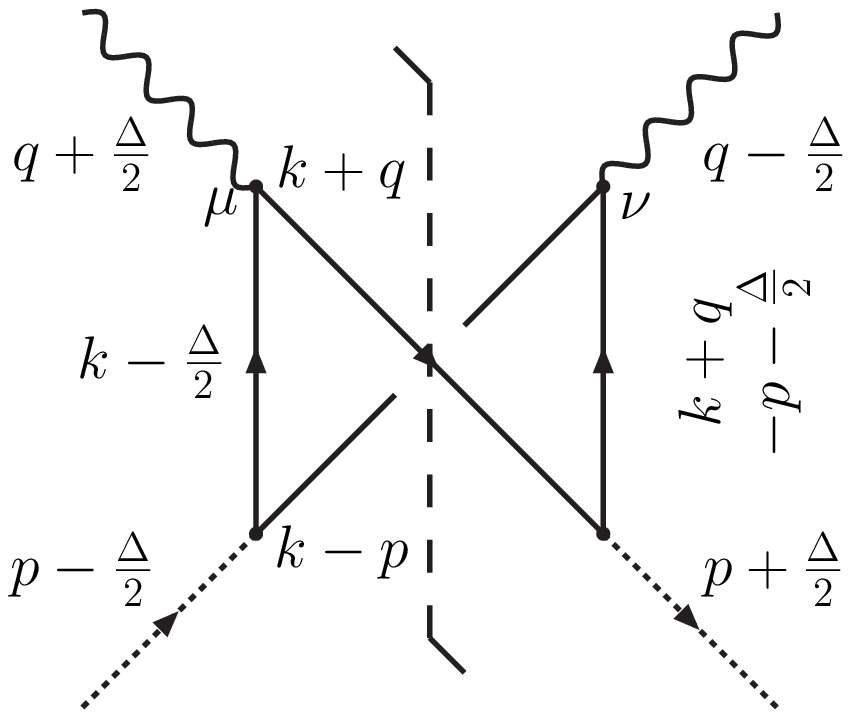,height=1.5cm}}
\caption[*]{The simplest diagrams contributing to the imaginary part of the 
amplitude for the scattering $\gamma^{\star} \pi \rightarrow 
\gamma^{\star}\pi$. 
Dashed lines
represent the discontinuity of the amplitudes, {\it i.e.} their imaginary parts.}
\label{fig:diagrams}
\end{figure}

We adopt the kinematics shown in Fig.~\ref{fig:diagrams}. We use the  Lorentz invariants $t = \Delta^2$,  $Q^2 = -q^2$, 
$x = Q^2/2p\cdot q$ and $\xi = \Delta\cdot q / 2p\cdot q$. The diagonal limit is characterised by $\xi=t=0$, 
the elastic limit by $\xi=0$, and the deeply virtual compton scattering (DVCS) limit by $\xi=-x$ for $t\ll Q^2$.

The  hadronic tensor $T_{\mu\nu} (q, p, \Delta)$ can be written, for a scalar or pseudoscalar target, as~\cite{Belitsky:2000vk}
\begin{eqnarray}
\label{eq:tmunudecomp}
&\textstyle{T_{\mu\nu} =
- {\cal P}_{\mu\sigma} g^{\sigma\tau} {\cal P}_{\tau\nu}
F_1
+ \frac{{\cal P}_{\mu\sigma} p^\sigma p^\tau {\cal P}_{\tau\nu}}{p \cdot q}
F_2
+ \frac{{\cal P}_{\mu\sigma} (p^\sigma (\Delta^\tau-2\xi p^\tau)
+  (\Delta^\sigma-2\xi p^\sigma) p^\tau) {\cal P}_{\tau\nu}}{2 p \cdot q}
F_3\nonumber}\\
&\textstyle{+ \frac{{\cal P}_{\mu\sigma}
(p^\sigma (\Delta^\tau-2\xi p^\tau)
-  (\Delta^\sigma-2\xi p^\sigma)p^\tau) {\cal P}_{\tau\nu}}{2 p \cdot q}
F_4
+ {\cal P}_{\mu\sigma}
(\Delta^\sigma-2\xi p^\sigma) (\Delta^\tau-2\xi p^\tau) {\cal P}_{\tau\nu}
F_5.}
\end{eqnarray}
Current conservation is guarenteed by means of the projector
${\cal P}_{\mu\nu}=g_{\mu\nu} - \frac{q_{2 \mu} q_{1 \nu}}{q_1 \cdot q_2}$,
where $q_1$ and $q_2$ are the momenta of the incoming and outgoing photons, respectively. The structure functions $F_i$ are functions of the invariant quantities $x$, $\xi$ and $t$. They are all even functions of $\xi$, except for $F_3$, which is odd.

\vspace*{-1cm}
\section{THE MODEL}
The  model introduced in our previous work~\cite{ours} includes massive pion and massive quark
fields and a pion-quark coupling described by the Lagrangian  interaction density
$ {\mathcal L}_{int}= i g (\overline{\psi}\mbox{  $\vec{\tau}$} \gamma_5 \psi) .
\mbox{  $\vec{\pi}$}$,
where $\psi$ is the quark field, $\vec{\pi}=(\pi^+,\pi^0,\pi^-)$ is the pion field and $\vec{\tau}$ is the isospin  operator.

At leading order in the loop expansion, four diagrams  contribute, see in Fig.~\ref{fig:diagrams}. 
We have evaluated their imaginary part, using the integration variables $\tau=k^2$, $k_{\rho}=|\vec{k}|$, $\phi$ 
and $\theta$,  the polar angles of $\vec{k}$ with respect
to  the direction of the incoming photon.   Actually, due to the discontinuity of the diagrams, indicated 
in Fig.~\ref{fig:diagrams}, we must integrate on $\tau$ and $\phi$ only. We do not give  the  
expressions here. They can be found in Ref.~\cite{our2} . However, we can sketch our procedure for 
imposing a finite size to the pion. The relative four-momentum squared of the quarks inside the pion is given by
\begin{eqnarray}
O^\pm&=&\left(2k-p\pm\frac{\Delta}{2}\right)^2 =
2\tau + 2 m_q^2 - m_{\pi}^2+\frac{t}{2}\pm 2k\cdot \Delta,
\end{eqnarray}
for pion-quark vertices like the ones in the first diagram of Fig.~\ref{fig:diagrams}. Note that
this can be rewritten as 
\begin{equation}
O^\pm=\left(k \pm \frac{\Delta}{2}\right)^2 +  2 m_q^2 - m_{\pi}^2.
\end{equation}
The first quantity in the r.h.s. being nothing but  the squared momentum transfer for the 
$\gamma^{\star} \pi \rightarrow q \overline{q}$ process, $O^\pm$ can be written as a function of the 
external variables for this process and of the masses. Similar expressions hold for other vertices. 
Generalizing the procedure of Ref.~\cite{ours}, we  require $|O^{\pm}| <
\Lambda^2$ either for one or  the other vertex of each diagram. Gauge invariance is therefore 
preserved by this cut-off, as it can be thought of as a constraint on the intermediate state cut 
lines. In practice, this is equivalent to requiring one of the two following conditions:
\begin{equation}
\textstyle{\tau <-\frac{\Lambda^2}{2}+ \frac{m_{\pi}^2}{2}-m_q^2-\frac{t}{4}+\left|k\cdot \Delta\right|,\ 
\tau > \frac{\Lambda^2}{2}- \frac{m_{\pi}^2}{2}+3m_q^2+\frac{t}{4}-\frac{Q^2}{x}
-\left|\frac{\xi Q^2}{x}+k\cdot \Delta\right|.}
\end{equation}
As explained in Ref.~\cite{ours}, owing to these conditions and for small $t$, the crossed diagrams
 are  suppressed
by a power $\Lambda^2/Q^2$, compared to  the box diagrams.

We keep the  coupling constant $g$ as in the diagonal case, where it was  determined by imposing that there are only two constituent quarks in the pion, or equivalently that the following relation
\begin{equation}
\int _{0}^{1}F_{1}dx=\frac{5}{18}
\end{equation}
holds, which makes $g$ dependent upon $Q^2$. It turns out that, with the cut-off, $g$ reaches an asymptotic value for $Q^2$ above 2 GeV$^2$. 
\section{Results for the structure functions}
\subsection{General features}
From the imaginary part of the total amplitude, the imaginary part of the five structure functions $F_i$ 
can be obtained by a projection on the corresponding tensors.
For any fixed value of $\xi$ not close 
to $\pm 1$, we recovered for $F_1$ and $F_2$ the same behaviour as in the diagonal case. 
We checked indeed that the diagonal limit is recovered  for $\xi=0$ and $t=0$. 
Furthermore the structure functions $F_3$, $F_4$, $F_5$ depend little on $\xi$ except 
when this variable is close to $\pm 1$.
In the particular case of DVCS, in the presence of finite-size effects, the value of $F_1$ gets significantly 
reduced, especially for small $x$,  as $|t|$ increases, whereas that effect is 
much less noticeable  without cut-off.
In the elastic case, the same suppression at small $x$ is observed, especially
when the cut-off is applied. 

The effect of the variation of $Q^2$ were also studied. As in the 
diagonal case~\cite{ours}, we can conclude that the details of the non-perturbative effects
cease to matter for $Q^2$ greater than 2 GeV$^2$, that is significantly larger than $\Lambda^2$. 

\subsection{High-$Q^2$ limit: new relations}

Having determined the 5 functions $F_i$'s in the context of our model, we 
shall now consider their behaviour at high $Q^2$. Expanding the ratios of 
$\frac{F_2}{F_1}$$,\frac{F_3}{F_1}$,$\frac{F_4}{F_1}$,$\frac{F_5}{F_1}$, 
we obtain the following asymptotic behaviour:
\begin{align}\label{eq:F_irel}
F_2&=2xF_1+{\cal O}(1/Q^2),&F_3=&\frac{2x\xi}{\xi^2-1}F_1+{\cal O}(1/Q^2),\\
F_4&=\frac{2x}{\xi^2-1}F_1+{\cal O}(1/Q^2),&F_5=&{\cal O}(1/Q^2). 
\end{align}
The first relation is similar (at leading order in $1/Q^2$ and with the replacement of $x$ by $x_B$) to the Callan-Gross relation 
between the diagonal structure functions $F_1$ and $F_2$, valid for spin one-half 
constituents in general. Except for $F_5$, which is small at large $Q^2$, these relations 
show that $F_2$, $F_3$ and $F_4$ are simply related to $F_1$ at leading order. 
They also clearly display and therefore confirm the 
symmetries of these functions. 
The fact of getting such simple relations between the $F_i$'s (at leading order) constitutes a remarkable
result of our model. Furthermore, we checked that the term ${\cal O}(1/Q^2)$
in the first relation is numerically quite small, even for moderate $Q^2$. 
One may wonder whether these results are  typical of our model, or more general. 

\section{Linking the $F_i$'s to ${H}$, ${H}^3$, and $\tilde {H}^3$}

Having at hand the five functions $F_i$'s that parametrise the amplitude for
$\gamma^\star\pi\to\gamma^\star\pi$, we can link them to the generalised parton distributions.
For this purpose, we make use of a tensorial expression coming from the twist-three analysis
 of the process, which singles out the twist-two ${\cal H}$ and the twist-three
 ${\cal H}^3,\tilde {\cal H}^3$ form factors. Following Ref.~\cite{Belitsky:2000vk}, we
write\footnote{Please note that Ref.~\cite{Belitsky:2000vk} uses ${\cal P}_{\nu\mu}$ instead of
${\cal P}_{\mu\nu}$ as projector.}:
\begin{eqnarray}\label{Tw3Tamplitude}
\textstyle{T_{\mu\nu} =
- {\cal P}_{\sigma\mu} g^{\sigma\tau} {\cal P}_{\nu\tau}
\frac{q \cdot V_1}{2p \cdot q}+ \left( {\cal P}_{\sigma\mu} p^\sigma  {\cal P}_{\nu\rho}
+ {\cal P}_{\rho\mu}  p^\sigma {\cal P}_{\nu\sigma} \right)
\frac{V_{2}^{\rho}}{p \cdot q}
- {\cal P}_{\sigma\mu} i\epsilon^{\sigma \tau q \rho} {\cal P}_{\nu\tau}
\frac{A_{1\, \rho}}{2p \cdot q}}.
\end{eqnarray}
where the $V_i$'s and $A_1$ read
\begin{eqnarray}\label{V1V2A1}
V_{1 \, \rho} &=& 2 p_\rho {\cal H} +
(\Delta_\rho -2 \xi p_\rho)  {\cal H}^3 
+ \mbox{twist 4},
A_{1 \, \rho} = \frac{i\epsilon_{\rho\Delta p q}}{p\cdot q}\;
{\tilde{\cal H}}^3,\\
V_{2 \, \rho} &=& x V_{1 \, \rho} - \frac{x}{2}
\frac{p_\rho}{p\cdot q} q\cdot V_{1} + \frac{i}{4}
\frac{\epsilon_{\rho\sigma\Delta q}}{p\cdot q} A_{1}^{\sigma}
+ \mbox{twist 4}.
\end{eqnarray}
In Ref.~\cite{Belitsky:2000vk}, gauge invariance of Eq.~(\ref{Tw3Tamplitude})  beyond 
the twist-three accuracy was in fact restored by hand, contrarily to 
the present calculation for which the amplitude is explicitly gauge invariant.

To relate the $F_i$'s to the $\cal H$'s, we project the amplitude~(\ref{Tw3Tamplitude})
onto the five projectors contained in Eq.~(\ref{eq:tmunudecomp}) and identify the results
with the $F_i$'s. Note that, in the neutral pion case, the imaginary part of
the form factors $\cal H$, ${\cal H}^3$ and $\tilde {\cal H}^3$ directly gives the GPD's 
$H$, $H^3$ and $\tilde H^3$ up to a factor $2\pi$. As we have
kept the off-shellness of the photons arbitrary, we in fact can relate the 
imaginary parts of $F_i$ to the GPD's for arbitrary $x$ and $\xi$ (up to ${\cal O}(1/Q^2)$ terms):
\begin{eqnarray}\hspace*{-0.6cm}
{F_1\over 2\pi}&=&{H}, \label{eq:FH1}  \hspace*{0.2cm}
{F_2\over 2\pi}=2x{H}, \label{eq:FH2}   \hspace*{0.2cm}
{F_3\over 2\pi}=\frac{2x}{x^2-\xi^2}\left({H}^3x^2+\tilde{H}^3\xi x-{H}\xi\right), \label{eq:FH3}    \\ 
{F_4\over 2\pi}&=&\frac{2x}{x^2-\xi^2}\left({H}^3\xi x+\tilde{H}^3x^2-{H}x\right), \label{eq:FH4} \hspace*{0.25cm}
{F_5\over 2\pi}={\cal O}(1/Q^2)\label{eq:FH5}
\end{eqnarray}

Replacing the $F_i$'s by the
expressions~(\ref{eq:F_irel}), we can write (up to ${\cal O}(1/Q^2)$ terms)
\begin{eqnarray}\label{eq:HH3rel}
\tilde{H}^3=\frac{(x-1)}{x(\xi^2-1)}H
\hbox{ and }
{H}^3=\frac{(x-1)\xi}{x(\xi^2-1)}H=\xi\tilde{H}^3.
\end{eqnarray}
As $F_1$ to $F_4$ can be written in term of only one of them, {\it e.g.} $F_1$, it
is not surprising that ${H}^3$ and ${\tilde H}^3$ are simply related
to ${H}$. Note that polynomiality of the Mellin moments of $H$, $H^3$ and $\tilde H^3$, together with 
Eqs.~(\ref{eq:HH3rel}), imply that
$H$ must be the product of  $\xi^2-1$ with a polynomial in $\xi$, $P_H$. Morevover, the fact that
$\tilde H_3$ is almost independent of
$\xi$ shows that $P_H$ is very close to a constant.

To convince ourselves that relations (\ref{eq:HH3rel}) are new, we have compared 
them to the Wandzura-Wilczek approximation \cite{WW}, given for the pion case
in \cite{Belitsky:2000vk,Polyakov}. First of all, it is well-known that
these relations are discontinuous at $\xi=\pm x$, which is not the
case for (\ref{eq:HH3rel}). Furthermore, we compared the results of
the Wandzura-Wilczek approximation with our results. We found that the two 
are numerically very different. Hence relations (\ref{eq:HH3rel}), derived
in an explicitly gauge-invariant model, do not come from "kinematical" 
twist corrections, but emerge from the dynamics of the spectator quark 
propagator and from finite-size effects.

\section*{Acknowledgements} 
The authors wish to thank M.~V. Polyakov for his useful comments.
This work has been performed in the frame of the ESOP collaboration 
(European Union contract HPRN-CT-2000-00130).

\end{document}